\documentclass[12pt]{article}
\usepackage{amsmath,amsfonts,epsfig}
\numberwithin{equation}{section}

\ifx\ltimes\undefined
\newcommand{\ltimes}{{\kern3pt\hbox{\vrule width 0.4pt height 5.30pt
depth .0pt}\kern-1.76pt\times\kern1pt}} \fi

\ifx\lrtimes\undefined
\newcommand{\rtimes}{{\kern1pt\times\kern-4.76pt\kern3pt\hbox{\vrule width 0.4pt height 5.30pt
depth .0pt}}} \fi

\def\Z {\mathbb{Z}}
\def\R {\mathbb{R}}

\def\ti{\tilde}

\def\a{\alpha}
\def\b{\beta}

\def\g{\gamma}

\def\s{\sigma}                                   

\def\D{\Delta}

\def\G{\Gamma}

\def\cG{{\cal G}}
\def\cX{{\cal X}}

\begin{document}

\begin{titlepage}

\thispagestyle{empty}
\begin{flushright}\footnotesize
\texttt{ArXiv}\\
\texttt{Imperial/TP/2007/CH/03}\\
\texttt{DESY-07-189}\\
\texttt{ZMP-HH/07-11}\\
\vspace{2.1cm}
\end{flushright}

\begin{center}

{\Large {\textbf{Gauge Symmetry, T-Duality and Doubled Geometry}}} \\

\vspace*{12mm}

{C.M. Hull$^{1}$ and R.A.~Reid-Edwards$^2$} \\
\vspace*{7mm}

{\em $^1$The Institute for Mathematical Sciences}\\
{\em Imperial College London} \\
{\em 53 Prince's Gate,  London SW7 2PG, U.K.} \\

\vspace*{4mm}

{\em $^1$The Blackett Laboratory, Imperial College London} \\
{\em Prince Consort Road, London SW7 2AZ, U.K.} \\

\vspace*{4mm}

{\em $^2$II. Institute fur Theoretische Physik }\\
{\em Universit\"{a}t Hamburg } \\ {\em DESY, Luruper Chaussee 149} \\
{\em D-22761 Hamburg, Germany} \\

\vspace*{12mm}

\end{center}

\begin{abstract}
\noindent String compactifications with T-duality twists are revisited and  the gauge algebra of the dimensionally reduced theories calculated.
These reductions can be viewed as string theory on T-fold backgrounds, and can be formulated in a \lq doubled space' in which  each circle is
supplemented by a T-dual circle to construct a geometry which is a doubled torus bundle over a circle. We discuss a conjectured extension to
include T-duality on the base circle, and propose the introduction of a dual base  coordinate, to give a doubled space which is locally the
group manifold of the gauge group. Special cases include those in which the doubled group is a Drinfel'd double. This gives a framework to
discuss backgrounds that are not even locally geometric.

\end{abstract}

\vfill

\noindent {c.hull@imperial.ac.uk\\ ronald.reid.edwards@desy.de}

\end{titlepage}

\newpage

\section{Introduction}

Two kinds of dimensional reduction of supergravities were proposed in the seminal paper of Scherk and Schwarz \cite{Scherk ``How To Get Masses
From Extra Dimensions''}, each involving a twist by a group. Each gives a lower dimensional supergravity, which typically is gauged, i.e. has a
non-Abelian Yang-Mills group. Recently   it was understood how to lift these dimensional reductions to the full supergravity, string theory or
M-theory \cite{Dabholkar ``Duality twists orbifolds and fluxes'',Hull ``Flux compactifications of string theory on twisted tori'',Hull ``Flux
compactifications of M-theory on twisted tori'',ReidEdwards ``Geometric and non-geometric compactifications of IIB supergravity''}. The key is
to show that each can arise from a compactification, so that the full massive spectrum is defined, including  Kaluza-Klein  modes, massive
string modes, wrapped branes etc. Understanding the compactification geometry is important in understanding the structure of the theory, and it
turns out that this is intimately related to the structure of the gauge group. Some of the reductions, those with T-duality twists, do not lift
to any compactification of supergravity. However, they can lift to non-geometric reductions of string theory on T-folds. Much remains to be
understood about such non-geometric backgrounds, and the aim here is to use the gauge algebra of the dimensionally reduced theory to gain some
insight into such reductions. In \cite{Hull ``A geometry for non-geometric string backgrounds'',Hull ``Doubled geometry and T-folds''}, it was
shown that string theory on a T-fold that looks like a $T^d$ bundle locally has a natural formulation on a bundle in which the torus fibres are
doubled to become $T^{2d}$. Our considerations here lead to a natural geometry in which all the dimensions are doubled, not just the fibres.

The first class of Scherk-Schwarz reductions look superficially like reductions on an $n$-torus, but twisted with the action of an
$n$-dimensional group  $G$. For this reason, they have become known, misleadingly, as twisted torus reductions. The reduction can be thought of
as choosing an internal space that is the group manifold for  $G$, which is typically non-compact, and then consistently truncating to fields independent of
the \lq internal' coordinates. In \cite{Hull ``Flux compactifications of string theory on twisted tori''}, it was shown that in most cases the
same theory can be obtained from compactification on a compact manifold which looks like the group manifold  locally. This requires the
existence of a discrete subgroup $\G\subset G$ such that $\cX=G/\G$ is compact (so that $\G$ is then a cocompact subgroup of $G$), in which case
the theory is simply compactified on $\cX=G/\G$. The Scherk-Schwarz ansatz  involves the expansion of the higher dimensional fields in terms of
a basis of globally defined one-forms $\{\sigma\}$. In order for the one-forms $\sigma$ to be globally defined on $\cX$, it is necessary that they are
invariant under the action of $\G$, so that the reduction ansatz is invariant under  $\G$. However, one of the consequences of the
reduction ansatz being invariant under  $\G$ is that the gauged supergravity contains little information about the global structure
of $\cX$.

If we include a constant flux for the $H$-field so that   $H\sim K_{mnp}\sigma^m\wedge\sigma^n\wedge\sigma^p+\dots $, the
supergravity, resulting from such a Scherk-Schwarz compactification on $\cX$, has gauge algebra
\begin{eqnarray}
[Z_m,Z_n]=f_{mn}{}^pZ_p+K_{mnp}X^p  \qquad  [X^m,Z_n]=f_{np}{}^mX^p \qquad  [X^m,X^n]=0
\end{eqnarray}
where $Z_m$ generate isometries of $\cX$ and $X^m$ generate  antisymmetric tensor transformations for the $B$-field (see \cite{Hull ``Flux compactifications of string theory on twisted tori'', Kaloper ``The O(dd) story of
massive supergravity''} for details). Here  $f_{mn}{}^p$ are the
structure constants of the group $G$.

The other type of Scherk-Schwarz reduction starts with  reduction on a torus $T^d$, so that the dimensionally reduced and truncated theory has a
continuous duality symmetry $K$.
This is followed by
 a further reduction on a circle with a duality twist, so that on going round the extra circle, the theory
comes back to itself transformed by a duality transformation. In the full string theory or M-theory, the duality symmetry is broken to a
discrete subgroup $K(\Z)$ \cite{Hull ``Unity of superstring dualities''} and for the reduction     to lift to string or M-theory, the monodromy
must be in $K(\Z)$ \cite{Hull ``Massive string theories from M-theory and F-theory''}.
 A subgroup $GL(d;\Z)$ of $K(\Z)$ acts geometrically as diffeomorphisms of the $d$-torus,
and if the monodromy $M$ is in $GL(d;\Z)$, the reduction can be thought of as compactification on a $d+1$ dimensional space $\cX$ which is a
$T^d$ bundle over a circle with monodromy $M$ \cite{Hull ``Flux compactifications of string theory on twisted tori'',Hull ``Massive string
theories from M-theory and F-theory''}. Moreover, as we shall review in section 2, such a space $\cX$ is in fact a twisted torus in the previous
sense, i.e. it is locally a group manifold, and is of the form $\cX=G/\G$ \cite{Hull ``Flux compactifications of string theory on twisted
tori''}. In the case of reduction of pure gravity, the group $G$ is precisely the gauge group of the reduced theory.

Here we will focus on the T-duality subgroup $O(d,d;\Z)\subseteq K(\Z)$. There is a geometric subgroup of $O(d,d;\Z)$ acting through torus
diffeomorphisms and integral shifts of the $B$-field. If the monodromy is not in this subgroup, it is not a  compactification, but can be
thought of as string theory with a non-geometric internal space, known as a T-fold  \cite{Hull ``A geometry for non-geometric string
backgrounds''}. However, as $O(d,d;\Z)\subset GL(2d;\Z)$, it has a natural action as diffeomorphisms of a \lq doubled torus' $T^{2d}$ and there
is a $T^{2d}$ bundle over a circle with such a monodromy. A formulation using a circle coordinate and its dual    arises naturally in the string
field theory for toroidal backgrounds \cite{Kugo ``Target space duality as a symmetry of string field theory''}. String theory
reduced in this way with a duality twist
  can be formulated as a sigma-model on
a bundle over a circle  whose fibres are  the
   doubled torus $T^{2d}$ \cite{Hull ``A geometry for
non-geometric string backgrounds''}.  The doubled formalism has the virtue that it provides a geometric interpretation to many nongeometric
backgrounds. The doubled torus has coordinates conjugate to both the $d$ momenta and the $d$ winding numbers. Different dual backgrounds arise
from choosing different polarisations or choices of $T^d\subset T^{2d}$, specifying the \lq real' spacetime slice of the doubled space.
T-duality acts to change the choice of polarisation, and T-folds arise when there is no global polarisation.

Such reductions with duality twists give theories with a gauge group of dimension $2(d+1)$, the same as it would be for reduction on an
untwisted $S^1\times T^d$. In that case, the group is $U(1)^{2(d+1)}$ with $U(1)^{d+1}$ from the natural geometric action on $S^1\times T^d$ and
another $U(1)^{d+1}$ from $B$-field gauge transformations. One might expect that   reductions with a duality twist would give a gauge group
containing $U(1)^{2d}$ associated with the $T^{d}$ fibres. As it happens, this is not the case, and we give
  a careful  derivation of the gauge algebra here. One of the aims of this paper is to explore the implications of the structure of the gauge algebra.

In the doubled formalism, the $T^d$ fibre   is doubled, and this raises the question of  whether the base $S^1$ might also be doubled. This
would be relevant for the issue of whether one can T-dualise over the base circle. As the geometry has non-trivial dependence on the $S^1$
coordinate $x$, there is no isometry on the circle so the usual formulations of T-duality do not apply. However, in \cite{Dabholkar
``Generalised T-duality and non-geometric backgrounds''} a generalisation of T-duality to such situations was proposed, in which dependence on
the circle coordinate $x$ is transformed under T-duality to dependence on the coordinate  of a dual circle, $\ti x$. In this context it is
natural to consider more general reductions involving independent duality twists over $x$ and $\tilde{x}$. Such backgrounds would not admit a
geometric description even locally.

Conventional considerations are insufficient to discuss the situation with non-trivial dependence on $\ti x$. Here, we identify a  $2(d+1)$
dimensional doubled geometry that extends the doubled torus bundle to include one other dimension, with coordinate $\ti x$, and which is the
group manifold of the gauge group, identified under a discrete subgroup to give a doubled twisted torus. This is the natural space to include
all possible dual backgrounds, including the ones involving the conjectured generalised T-duality on the base $S^1$. The full gauge group
contains generators acting geometrically  on the original space  and ones acting as $B$-field gauge transformations, while in this  doubled
picture, all arise geometrically.

The canonical example of the types of string background discussed above is given by a sequence of T-dualities starting from the
three-dimensional nilmanifold. This nilmanifold is a $T^2$ bundle over $S^1$ with monodromy given by a parabolic   element of $SL(2;\Z)$.
T-duality interchanges various quantities referred to as generalised fluxes in \cite{Shelton ``Nongeometric flux compactifications''} and called
$f$-flux, $Q$-flux and $R$-flux in \cite{Shelton ``Nongeometric flux compactifications''}.
 As will be reviewed in the next section,
the nilmanifold may be thought of as a twisted torus characterized by the structure constant (or `geometric flux') $f_{xz}{}^y=m\in \Z$
\cite{Hull ``Massive string theories from M-theory and F-theory'', Hull ``Flux compactifications of string theory on twisted tori''}. The Buscher rules can be applied fibrewise to dualise along the $T^2$ fibre directions. Dualising along the $y$ direction gives a $T^3$ with $H$-flux
$K_{xyz}=m$, whilst dualising along the $z$ direction gives a T-fold, characterized by the `flux' $Q_x{}^{zy}=m$. It was   conjectured in
\cite{Dabholkar ``Generalised T-duality and non-geometric backgrounds''} that a further T-duality along the $x$-direction   gives rise to a
background constructed as a $T^2$ fibration over the dual coordinate $\tilde{x}$ with a T-duality twist. This conjectured background is
characterized by the nongeometric flux $R^{xyz}=m$ (or `R-flux'). The duality sequence may be summarized as \cite{Shelton ``Nongeometric flux
compactifications''}
\begin{equation}
K_{xyz}\rightarrow f_{xz}{}^y\rightarrow Q_{x}{}^{yz}\rightarrow R^{xyz}
\end{equation}
Whilst the doubled formalism has been successfully employed to give a geometric description of the T-fold, such an understanding of the
backgrounds with $R$-flux has not been forthcoming. It is the aim of this paper to shed some light on the group theoretic and geometric
structures which underly the duality sequence above. In particular we shall see that a knowledge of the gauge algebra of the compactified theory
suggests a natural  local structure for a doubled internal space.

A key objective is to understand how to lift a general gauged supergravity to superstring theory. The   structure constants of the gauge algebra
can be thought of as arising from the various types of flux, and so this is a question of understanding backgrounds with $f,H,Q$ or $R$ fluxes, and in particular
the non-geometric ones with $Q$ or $R$ flux.

The plan of this paper is as follows: In the next section we review the relationship between duality twist backgrounds with geometric monodromy
and  twisted tori. Section 3 will consider the general $O(d,d)$-twisted reduction. In section 4 the Yang-Mills gauge symmetries of this theory
will be studied. Section 5 describes the doubled geometry of the backgrounds considered here.

\section{Reductions  with a Geometric Duality Twist}

In this section we review the Scherk-Schwarz reduction with a geometric twist \cite{Hull ``Flux compactifications of string theory on twisted
tori''} and its  relation to a reduction on a twisted torus of the form $G/\G$.

Consider a $D+d+1$ dimensional field theory. We reduce the theory on a $d$-dimensional torus $T^d$, with real coordinates $z^a\sim z^a+1$ where
$a=1,2...d$. This produces a theory in $D+1$ dimensions with scalar fields that include those in the coset $GL(d;\R)/SO(d)$ arising from the
torus moduli. Truncating to the $z^a$ independent zero mode sector gives a theory that has a rigid $GL(d;\R)$ symmetry, while in the full Kaluza-Klein
theory this is broken to $GL(d;\Z)$ -- the mapping class group of the $T^d$. Let
\begin{equation}
ds^2=\hat{G}_{ab}dz^a dz^b
\end{equation}
 be the metric on the
$d$-torus. The symmetric matrix  $\hat{G}_{ab}$ parameterises  the   moduli space   $GL(d;\R)/SO(d)$. There is a natural action of $GL(d;\R)$
 on the metric and coordinates $z^a$ in which
\begin{equation}\label{coset symmetries}
\hat{G}_{ab}\rightarrow (U^t)_a{}^c\hat{G}_{cd}U^d{}_b    \qquad z^a\rightarrow (U^{-1})^a{}_b z^b
\end{equation}
where $U^b{}_a\in GL(d,\R)$.  We now truncate to a massless $D+1$ dimensional field theory and consider reduction on a further circle. In the
twisted reduction, dependence on the circle coordinate $x$ is introduced through a $GL(d;\R)$ transformation $U=\gamma(x)$ where
$\gamma(x)=\exp\left(Nx\right)$ and $N^a{}_b$ is some matrix in the Lie algebra of $GL(d;\R)$. This defines the $x$-dependence of the torus
moduli through
\begin{equation}
G(x)_{ab}=(\gamma (x)^t)_a{}^c\hat G _{cd}\gamma (x)^d{}_a
\end{equation}
for some arbitrary choice $\hat G _{ab}$. The monodromy round the circle  $x\sim x+1$ is   $e^N\in GL(d;\R)$. The truncation of all Kaluza-Klein
modes gives the Scherk-Schwarz reduction \cite{Hull ``Flux compactifications of string theory on twisted tori''}. A necessary condition for this
to lift to a compactification of the original $D+d+1$ dimensional theory, keeping all Kaluza-Klein modes,  is that the monodromy is in
$GL(d;\Z)$, which puts strong constraints on the choice of $N$ \cite{Hull ``Gauged D = 9 supergravities and Scherk-Schwarz reduction''}.
Assuming $e^N\in GL(d;\Z)$,
  the twisted reduction is equivalent to the
reduction on a $T^d$ bundle over $S^1$ with metric
\begin{equation} \label{metricbun}
ds_{d+1}^2=dx^2+G(x)_{ab}dz^adz^b=(\sigma^x)^2+\hat G_{ab}\sigma^a\sigma^b
\end{equation}
where
\begin{equation} \label{forms}
\sigma^x=dx \qquad \sigma(x)^a=\gamma(x)^a{}_bdz^b
\end{equation}

We now consider the group structure of this space. The forms (\ref{forms}) are globally defined on the torus bundle, and satisfy
\begin{equation}\label{parellelisability}
d\sigma^x=0 \qquad  d\sigma^a-N^a{}_b\sigma^x\wedge\sigma^b=0
\end{equation}
This $d+1$ dimensional space is then parallelisable, and locally looks like a group manifold $G$ with left-invariant Maurer-Cartan forms $\s$ associated with the
Lie algebra
\begin{equation}\label{duality twist algebra}
[t_x,t_a]=-N_a{}^bt_b, \qquad [t_a,t_b]=0
\end{equation}
This algebra can be represented by the $(d+1)\times(d+1)$ matrices
\begin{equation}
t_x=\left(\begin{array}{cc}-N^a{}_b & 0 \\ 0 & 0
\end{array}\right)  \qquad  t_a=\left(\begin{array}{cc}0 & e_a \\ 0 & 0
\end{array}\right)
\end{equation}
where $e_a$ is the $d$-dimensional column vector with a 1 in the a'th  position and zeros everywhere else. A representation of this Lie
algebra  is given by
\begin{eqnarray}
Z_x=\partial_x-N_a{}^bz^a\partial_b   \qquad Z_a=\partial_a
\end{eqnarray}
These vector fields are invariant under the left action of the group and are dual to the one forms $\sigma$. Coordinates $x, z^a$ can be
introduced locally for the group manifold, with the group element
 $g=g(x,z^a)\in G$
 given by
 \begin{equation}\label{groupG}
g=\left(\begin{array}{cc}\gamma^{-1}(x) & z \\ 0 & 1
\end{array}\right)
\end{equation}
Then the left-invariant Maurer-Cartan forms are given by
\begin{equation}
g^{-1}dg=\left(\begin{array}{cc}-N^a{}_b\sigma^x & \sigma^a \\
0 & 0
\end{array}\right)=\sigma^mt_m
\end{equation}
in agreement with (\ref{forms}), where $m=1,2,..d+1$. The $T^d$ bundle over $S^1$ with metric (\ref{metricbun})
 has the same local geometry as this group manifold.

The torus bundle over a circle is obtained from the
   compactification of this non-compact group manifold   under the  identification by a discrete subgroup $\G$, acting from the left.
The left action of
\begin{equation}
h(\alpha,\beta^a)=\left(\begin{array}{cc} {\g}^{-1}(\alpha)^a{}_b & \beta^a \\ 0 & 1 \end{array}\right)
\end{equation}
is
\begin{equation}
g(x,z^a)\to h(\alpha,\beta^a)\cdot g(x,z^a)
\end{equation}
and acts on the coordinates through
\begin{equation}
x\to x+\alpha  \qquad  z^a\to (e^{-N\alpha})^a{}_b z^b+\beta^a
\end{equation}
with $\a ,\b^a \in \Z$ and  form a discrete subgroup $\G=\{h(\alpha,\beta^a)\in G\mid \a ,\b^a \in \Z\}$ and we can identify  the group manifold
$G$ under $\Gamma$. This gives a compact space $G/\Gamma$, and is identical to the torus bundle over a circle with metric (\ref{metricbun})
\cite{Hull ``Flux compactifications of string theory on twisted tori''}.

\section{Reduction with an $O(d,d)$ twist}

We now turn to the duality-twisted reduction of theories with a metric and $B$-field, and we will be particularly interested in the cases that
arise from string theory. Consider the theory in $D+d+1$ dimensional spacetime with Lagrangian
\begin{equation}
\label{D+d+1 lagrangian} {\cal L}_{D+d+1}=e^{-\widehat{\Phi}}\left( \widehat{R}*1-d\widehat{\Phi} \wedge *d\widehat{\Phi} -
\frac{1}{2}\widehat{G}_{(3)}\wedge *\widehat{G}_{(3)} \right)
\end{equation}
where $\widehat{G}_{(3)}=d\widehat{B}_{(2)}$. The compactification on $T^d$, using the standard Kaluza-Klein ansatz  gives \cite{Maharana
``Noncompact symmetries in string theory''}  a massless field theory with gauge group $U(1)^{2d}\subset O(d,d)$ and a manifestly $O(d,d)$ invariant
Lagrangian
\begin{eqnarray}\label{O(d,d) Lagrangian}
{\cal L}_{D+1}&=&e^{-\phi}\left(R*1+*d\phi\wedge d\phi+\frac{1}{2}*G_{(3)}\wedge G_{(3)}+\frac{1}{4}*d{\cal M}^{AB}\wedge d{\cal M}_{AB}\right.
\nonumber\\ &&\left.-\frac{1}{2}{\cal M}_{AB}*{\cal F}^A\wedge{\cal F}^B\right)
\end{eqnarray}
The details of this reduction are given in \cite{Hull ``Flux compactifications of string theory on twisted tori'',Kaloper ``The O(dd) story of
massive supergravity'',Maharana ``Noncompact symmetries in string theory''} and the conventions of \cite{Hull ``Flux compactifications of string
theory on twisted tori''} have been used. The scalar coset space $O(d,d)/O(d)\times O(d)$ is parameterised by a symmetric metric on this coset
${\cal M}_{AB}$, satisfying the constraint
\begin{equation}\label{M=LML}
{\cal M}_{AB}=L_{AC}({\cal M}^{-1})^{CD}L_{BD}
\end{equation}
where $L_{AB}$ is the constant $O(d,d)$ invariant metric, which is used to raise and lower the indices $A,B=1,...,2d$.

We   then reduce on a further circle, with coordinate $x\sim x+1$, with an $O(d,d)$ duality twist. The twist is specified by $N^A{}_B$,   a
matrix representation of an element of the Lie algebra of $O(d,d)$, and the $x$-dependence is given in terms of an $O(d,d)$ transformation $\exp
(Nx)$. The theory has a Yang-Mills sector with a gauge group with structure constants $t_{MN}{}^P$
that will be discussed in the next subsection.
The reduced theory may be written in a manifestly $O(d+1,d+1)$ covariant way
\begin{eqnarray}\label{O(d+1,d+1) Lagrangian}
{\cal L}_D&=&e^{-\varphi}\left(R*1+*d\varphi\wedge d\varphi+\frac{1}{2}*{\cal H}_{(3)}\wedge {\cal H}_{(3)}+\frac{1}{4}*D{\cal M}_{MN}\wedge
D{\cal M}^{MN}\right. \nonumber\\ &&\left.-\frac{1}{2}{\cal M}_{MN}*{\cal F}^M\wedge{\cal F}^N\right)+V*1
\end{eqnarray}
with $O(d+1,d+1)$ indices $M,N=1,...,2(d+1)$ that are raised and lowered using the  constant $O(d+1,d+1)$ invariant  metric $L_{MN}$. The
two-form field strengths ${\cal F}^M$ are written in terms of   connection one-forms ${\cal A}^M$ and the three-form ${\cal H}_{(3)}$ is written
in terms of the two-form potential $C_{(2)}$
\begin{equation}
{\cal F}^M=d{\cal A}^M+\frac{1}{2}t_{NP}{}^M{\cal A}^N\wedge{\cal A}^P  \qquad  {\cal H}_{(3)}=dC_{(2)}+\frac{1}{2}L_{MN}{\cal A}^M\wedge d{\cal
A}^N-\frac{1}{3}t_{MNP}{\cal A}^M\wedge{\cal A}^N\wedge{\cal A}^P
\end{equation}
where
$t_{MNP}=t_{MN}{}^QL_{PQ}$.
The scalars ${\cal M}_{MN}$ take values in the coset space $O(d+1,d+1)/O(d+1)\times O(d+1)$ and satisfy a constraint similar to (\ref{M=LML}).
The scalar potential is
\begin{equation}
V=e^{-\varphi}\left(\frac{1}{4}{\cal M}^{MQ}L^{NT}L^{PS}t_{MNP}t_{QTS}- \frac{1}{12}{\cal M}^{MQ}{\cal M}^{NT}{\cal M}^{PS}t_{MNP}t_{QTS}\right)
\end{equation}
Details of the reduction and the explicit forms of the potential and scalars in terms of the $D+1$ dimensional fields are given in   appendix A.

\noindent\textbf{Gauge Symmetry}

The $D+1$ dimensional theory (\ref{O(d,d) Lagrangian}) obtained from conventional reduction on $T^d$ has $U(1)^{2d}\subset O(d,d)$ gauge
symmetry. $U(1)^d$ comes from the isometry group of the internal $T^d$ and a further $U(1)^d$ comes from the antisymmetric tensor
transformations of the $B$-field. The generators of this gauge group $T_A$ ($A=1,...2d$) satisfy $[T_A,T_B]=0$. The duality twist reduction on a
further circle with coordinate $x$ to $D$ dimensions gauges a non-Abelian subgroup $\cG\subset O(d+1,d+1)$ given by the algebra
\begin{eqnarray}\label{O(d,d+16) Lie algebra}
\left[Z_x,T_A\right]&=&-N^B{}_AT_B   \qquad\qquad  \left[T_A,T_B\right]=-N_{AB}X^x
\end{eqnarray}
where $Z_x$ generates   shifts in the circle coordinate $x$ and $X^x$ is the generator of antisymmetric tensor transformations of the
$B$-field component with one leg along the $x$-direction and one leg in the external spacetime. All other commutators vanish.
Here
 \begin{equation}
N_{AB}=L_{[A|C}N^C{}_{|B]} =- N_{BA}
\end{equation}
The antisymmetry of $N_{AB}$ follows from the requirement that $N^A{}_B$ be a generator of $O(d,d)$. Note that the algebra satisfied by the
generators $T_A$ which can be associated with the action on $T^d$ has been deformed and is no longer Abelian.

The generators
\begin{eqnarray}
T_M=\left(%
\begin{array}{c}
  Z_x   \\
  X^x \\
  T_A \\
\end{array}%
\right)
\end{eqnarray}
satisfy a Lie algebra $[T_M,T_N]=t_{MN}{}^PT_P$ where $t_{MN}{}^P$ are the structure constants given
by
\begin{equation}
t _{xB}{}^A=-N^A{}_B, \qquad t_{x[AB]}=-N_{AB}
\end{equation}
The
derivation of this algebra is given in   appendix B.

The gauging introduces a deformation of the ungauged theory involving the  $t_{MN}{}^P$, which  breaks the rigid $O(d+1,d+1)$ symmetry of the ungauged theory to the
subgroup preserving the $t_{MN}{}^P$. However, the theory becomes formally invariant under $O(d+1,d+1)$ if the structure constants $t_{MN}{}^P$
are taken to transform covariantly under $O(d+1,d+1)$ \cite{Hull ``Flux compactifications of string theory on twisted tori'', Kaloper ``The
O(dd) story of massive supergravity''}.

\section{Lifting to String Theory}

The discussion so far has used the framework of conventional field theory. In this section we discuss the  lift of these results
to string theory.

The   $T_A$ generators consist of the $Z_a$  which generate the $U(1)^d$ action on the $T^d$ fibre and the
 $X^a$ which generate antisymmetric tensor transformations for the
$B$-field components with one leg on the $T^d$ and the other in the external spacetime, so that
\begin{eqnarray}
T_A=\left(%
\begin{array}{c}
  Z_a \\
  X^a \\
\end{array}%
\right)
\end{eqnarray}
The twist matrix then decomposes as (using $N_{AB}=-N_{BA}$)
\begin{eqnarray}
N^A{}_B=\left(%
\begin{array}{cc}
  f_{xa}{}^b &   Q_x{}^{ab}  \\
 K_{xab}   & -f_{xb}{}^a \\
\end{array}%
\right)
\end{eqnarray}
for some antisymmetric $Q_x{}^{ab}=-Q_x{}^{ba}, K_{xab}   =- K_{xba}$. The gauge algebra is then
\begin{eqnarray}
[Z_x,Z_a]=f_{xa}{}^bZ_b+K_{xab}X^b   \qquad  [Z_x,X^a]=-f_{xb}{}^aX^b+Q_x{}^{ab}Z_b
\end{eqnarray}
\begin{eqnarray}
[Z_a,Z_b ]=K_{xab}X^x  \qquad [X^a,Z_b]=-f_{xb}{}^aX^x \qquad  [X^a,X^b]=Q_x{}^{ab}X^x
\end{eqnarray}
with all other commutators vanishing.

If $Q_x{}^{ab}=0$, then the twist is geometric, consisting of a $GL(d;\Z)$ twist with $f_{xa}{}^b=N_a{}^b$ acting as a diffeormorphism of the
$T^d$ fibres generated by $N_a{}^b$ together with a B-shift acting on the fibre components of $B$, $B_{ab}\to B_{ab}+\a K_{xab}$. This is
equivalent to the compactification with flux $K$ on a $T^d$ torus bundle over a circle and, as reviewed in section 2, this is a twisted torus
$G/\Gamma$ where $G$ is the $d+1 $ dimensional group of matrices of the form (\ref{groupG}). The $K_{xab}$ gives a constant flux $K_{xab}
\sigma^x\wedge \sigma^a\wedge \sigma^b$ on the  twisted torus. These backgrounds have been studied extensively in \cite{Hull ``Flux
compactifications of string theory on twisted tori'',Hull ``Flux compactifications of M-theory on twisted tori'', ReidEdwards ``Geometric and
non-geometric compactifications of IIB supergravity'',Kaloper ``The O(dd) story of massive supergravity''}.

If  $Q_x{}^{ab}\neq 0$, the twist is non-geometric and involves T-dualities, so that the resulting background is a T-fold. As was shown in
\cite{Dabholkar ``Generalised T-duality and non-geometric backgrounds'',Shelton ``Nongeometric flux compactifications'', Hull ``Global Aspects
of T-Duality Gauged Sigma Models and T-Folds''} backgrounds with just one of these three structure constants switched on can be related to one
another by T-duality so that  T-duality is expected to be a symmetry of the full string theory which identifies certain $H$-flux, twisted torus
and T-fold compactifications as equivalent descriptions of the same physics.

The twist means that there  is no isometry on the  final circle acting to shift the  coordinate $x$. Nonetheless, there is some evidence that
there should still be a T-duality on this circle   \cite{ Dabholkar ``Generalised T-duality and non-geometric backgrounds''}
 that
 exchanges $Z_x$ with $X^x$ and would act on the structure constants as
\begin{equation}
K_{xab}\rightarrow f_{ab}{}^x    \qquad    f_{xa}{}^b\rightarrow Q_a{}^{xb}    \qquad  Q_x{}^{ab}\rightarrow R^{xab}
\end{equation}
to give the algebra
\begin{eqnarray}
[X^x,Z_a]=Q_{a}{}^{xb}Z_b+f_{ab}{}^xX^b   \qquad  [X^x,X^a]=-Q^{xa}{}_{b}X^b+R^{xab}Z_b
\end{eqnarray}
\begin{eqnarray}
[Z_a,Z_b ]=f_{ab}{}^xZ_x  \qquad [X^a,Z_b]=-Q^{xa}{}_{b}Z_x \qquad  [X^a,X^b]=R^{xab}Z_x
\end{eqnarray}
It was conjectured in \cite{Dabholkar ``Generalised T-duality and non-geometric backgrounds''} that the structure constant $R^{xab}$
(`$R$-flux') corresponds to a background constructed with a   twist over a dual circle $\widetilde{S}^1$ (with coordinate $\tilde x$ conjugate
to the winding number). In the next section we propose a geometric interpretation for all of these backgrounds and show that it supports this
interpretation of the $R$-flux.

\section{Doubled Geometry }

In section 2 we considered the case of a twisted reduction which has a simple geometric interpretation as a compactification on a $T^d$ bundle
over $S^1$ in which the torus moduli have monodromy in $GL(d;\Z)$ round the base circle. The internal space is a twisted torus, or group
manifold identified under a discrete subgroup. Including a monodromy that shifts the B-field corresponds to adding an $H$-flux to the twisted
torus. We now turn to the geometric interpretation of the T-duality twisted reductions of section 3.

\noindent\textbf{The Doubled Torus}

For the general (nongeometric) case a geometric approach has been given by the doubled torus formalism of \cite{Hull ``A geometry for
non-geometric string backgrounds''}. The $O(d,d;\Z)$ duality twist acts non-geometricaly on the torus $T^d$ (mixing the metric and B-field, for
example) but as $O(d,d;\Z)\subset GL(2d;\Z)$, it has a natural action as diffeormorphisms  of a doubled torus $T^{2d}$. There is then a $T^{2d}$
bundle over a circle with twist generated by $N^A{}_B$ constructed as in section 2. Such a doubled torus arises naturally in string theory, with
the original $d$ coordinates $z^a$ on $T^d$ conjugate to the momenta and an additional $d$ coordinates $\ti z_a$ conjugate to the winding
numbers on the original $T^d$. The $O(d,d;\Z)$ duality  group acts naturally on the periodic doubled coordinates
$\mathbb{X}^A=(z^a,\tilde{z}_a)$. It was shown in \cite{Hull ``A geometry for non-geometric string backgrounds'',Hull ``Doubled geometry and
T-folds''} that string theory compactified in this way could be formulated in terms of a sigma model with target given by this doubled torus
bundle. In this formalism, T-duality is a manifest symmetry, and the conventional formalism is recovered on choosing a polarisation, i.e. a $T^d
\subset T^{2d}$ which is to be regarded as the real spacetime torus. T-duality can be viewed as acting to change the choice of $T^d \subset
T^{2d}$, changing the geometry to a dual one. All   dual geometries are encoded in the doubled torus bundle. For a geometric background,  a
global polarisation can be chosen, but for T-folds the best one can do is choose a polarisation locally. The T-duality transition functions then
give the changes in polarisation from patch to  patch.

A doubled torus bundle over a circle is a twisted torus $G'/\Gamma'$, as in section 2. Simply applying the construction of section 2 to the
doubled torus gives a background in which the group $G'$ has generators $T_A, Z_x$ satisfying the algebra
\begin{equation} \label{sdfadfg}
\left[Z_x,T_A\right]=-N^B{}_AT_B    \qquad  \left[T_A,T_B\right]=0
\end{equation}
acting on the coordinates $(x, \mathbb{X}^A)$ as
\begin{equation}
Z_x=\partial_x+N^A{}_B\mathbb{X}^B\partial_A   \qquad T_A=\partial_A
\end{equation}
This algebra does not capture the full gauge algebra (\ref{O(d,d+16) Lie algebra}).
It is not a subalgebra, but it is the algebra acting on the sector in which $X^x$ acts trivially.
 In order to give a full geometric interpretation to the gauge algebra
$(\ref{O(d,d+16) Lie algebra})$ we need to extend the doubled torus construction.

\noindent\textbf{The Doubled Group}

The doubled torus formalism in which the fibres are doubled is useful for discussing T-duality on the fibres and  the various T-dual spaces arise as
different polarisations of the doubled torus bundle. If, as suggested in \cite{Dabholkar ``Generalised T-duality and non-geometric
backgrounds''}, one can also T-dualise on the base circle with coordinate $x$, it is natural to ask whether there is a doubled space that would
include a  T-dual circle to the base circle so that all T-dual spaces are incorporated as different $d+1$ dimensional polarisations of a
$2(d+1)$ dimensional space $\cX$. In each polarisation, half of the gauge group generators (the ones we have denoted $Z$) might be expected to
act geometrically on the $d+1$ dimensional space (in the simplest cases, these generate diffeomorphisms of the space). For this to apply for any
polarisation, it is natural to expect that the full gauge group (generated by the $Z$'s and $X$'s with Lie algebra (\ref{O(d,d+16) Lie
algebra})) should act on the doubled space. Comparison with the twisted torus construction suggests then that the doubled space should be
locally a group manifold $\cG$, with Lie algebra (\ref{O(d,d+16) Lie algebra}), identified under a discrete subgroup.

As in the discussion of the twisted torus geometry, one can represent the Lie algebra (\ref{O(d,d+16) Lie algebra}) in terms of the $2(d+1)$
coordinates $(x, \ti x, \mathbb{X}^A)$ of $\cG$, where $\mathbb{X}^A$ are the coordinates on the doubled torus fibre $T^{2d}$, as
\begin{eqnarray}\label{left generators}
\begin{array}{ll}
Z_x=\partial_x+N^A{}_B\mathbb{X}^B\partial_A   \qquad X^x=\partial_{\ti x} \qquad T_A=\partial_A-\frac{1}{2}N_{AB}\mathbb{X}^B\partial_{\ti x}
\end{array}
\end{eqnarray}
Then $X^x$ acts as translation in the new coordinate $\ti x$ and so acts trivially on fields that are independent of $\ti x$, so on such fields
the algebra (\ref{sdfadfg}) is realised and in this sector the doubled torus bundle gives a full geometric representation of the structure.
However, the doubled group gives a non-trivial extension to the general case with $\ti x $ dependence.

 The one forms dual
to these vector fields   satisfy the Maurer-Cartan equations
\begin{equation}
d{\cal P}^A-N^A{}_BP^x\wedge {\cal P}^B=0   \qquad \qquad   dQ_x-\frac{1}{2}N_{AB}{\cal P}^A\wedge{\cal P}^B=0  \qquad\qquad    dP^x=0
\end{equation}
which are solved by\footnote{The one-forms (\ref{doubled forms}) are dual to the vectors
\begin{equation}
Z_x=\partial_x  \qquad  X^x=\partial_{\tilde{x}}    \qquad
T_A=\left(e^{-Nx}\right)_A{}^B\left(\partial_B-\frac{1}{2}N_{BC}\mathbb{X}^C\partial_{\tilde{x}}\right)
\end{equation}
By a coordinate redefinition $\mathbb{X}^A\rightarrow\left(e^{Nx}\right)^A{}_B\mathbb{X}^B$, these vector fields are brought to the simpler form
(\ref{left generators}).}
\begin{equation}\label{doubled forms}
{\cal P}^A=\left(e^{Nx}\right)^A{}_Bd\mathbb{X}^B   \qquad \qquad   Q_x=d\tilde{x}+\frac{1}{2}N_{AB}\mathbb{X}^Ad\mathbb{X}^B  \qquad\qquad
P^x=dx
\end{equation}
This   is a doubling  of the geometry given
for the twisted torus in section 2, and the one-forms (\ref{doubled forms}) are the doubling of the one-forms (\ref{forms}).  The ${\cal P}^A$
and $P^x$ together describe the doubled torus fibred over $S^1$, but  a fully geometric interpretation of the gauge algebra requires a $2d+2$
dimensional space $\cG$ into which the doubled torus fibration is non-trivially embedded. It is useful to define the coordinates
$\mathbb{X}^I=\left(x,\tilde{x},\mathbb{X}^A\right)$ on the doubled group and ${\cal P}^M={\cal P}^M{}_Id\mathbb{X}^I$ as the one forms on $\cG$ satisfying  the  Maurer-Cartan equations
\begin{eqnarray}
d{\cal P}^M+\frac{1}{2}t_{NP}{}^M{\cal P}^N\wedge{\cal P}^P=0
\end{eqnarray}
where $t _{xB}{}^A=-N^A{}_B$ and $t_{x[AB]}=-N_{AB}$.

\noindent\textbf{T-Duality and $R$-Flux}

In the doubled torus picture, choosing a polarisation corresponded to choosing a  maximally isotropic subspace (null with respect to the
constant $O(d,d)$ metric $L_{AB}$) $T^d\subset T^{2d}$ as the geometric space with coordinates $z^a$ (and geometric generators $Z_a$) and the
complement $\widetilde{T}^d$, with coordinates $\tilde{z}_a$ (and generators $X^a$). As  $ \cG \subset O(d+1,d+1) $, it preserves the
$O(d+1,d+1) $ invariant metric $L_{MN}$, and an isotropic subspace of $\cG$ is one which is completely null with respect to this metric. In the
doubled group case, a choice of polarisation can be given by choosing a maximally isotropic subgroup $G\subset \cG$ (i.e. one whose generators
are all null with respect to $L_{MN}$).
 The geometry of the conventional sigma-model is given locally by $G$.
 In some cases, the complement of $G$ will also be a group $\widetilde{G}$, and this defines a dual polarisation.
  For example,
if the gauge algebra is a Poisson-Lie algebra, it takes the form
\begin{equation}
[Z_m,Z_n]=f_{mn}{}^pZ_p \qquad  [Z_m,X^n]=f_{mp}{}^nX^p+Q_m{}^{np}Z_p   \qquad  [X^m,X^n]=Q_p{}^{mn}X^p
\end{equation}
and has two maximally isotropic sub-algebras; one generated by $Z_m$ and the other by $X^m$, where $m=(a,x)$ . These generate two subgroups $G$
and $\widetilde{G}$ and either can be used to define a physical subspace, giving two, locally geometric, string backgrounds.

For general groups $\cG$, however, it may be the case that there is no suitable subgroup that can be used to define the desired polarisation, so
that one has to use the doubled picture and cannot eliminate half of the coordinates even locally. This is precisely the situation that leads to
the locally nongeometric $R$-space, which we now discuss.

In the doubled group picture, one might expect a generalisation of T-duality which acts on all the coordinates $\mathbb{X}^I$. This allows us to
consider the possibility of choosing either $x$ or its dual $\tilde{x}$ as the geometric coordinate in a polarisation. This is to be contrasted
with the doubled torus picture which, a priori, fixes $x$ to be the geometric coordinate and only doubles the fibres.

Acting with the conjectured T-duality on the algebra (\ref{O(d,d+16) Lie algebra}) which exchanges $x$ and $\tilde{x}$ produces the gauge
algebra
\begin{eqnarray}
\left[X^x,T_A\right]&=&-N^B{}_AT_B   \qquad\qquad  \left[T_A,T_B\right]=-N_{AB}Z_x
\end{eqnarray}
which has corresponding one-forms
\begin{equation}
{\cal P}^A=\left(e^{N\tilde{x}}\right)^A{}_Bd\mathbb{X}^B   \qquad \qquad   P^x=dx+\frac{1}{2}N_{AB}\mathbb{X}^Ad\mathbb{X}^B \qquad\qquad
Q_x=d\tilde{x}
\end{equation}
which is an $O(d,d)$ twist over the dual coordinate $\tilde{x}$ as conjectured in \cite{Dabholkar ``Generalised T-duality and non-geometric
backgrounds''}. This has non-trivial dependence on the dual coordinates $\ti x$, so cannot be interpreted as a conventional background even
locally. This space is a $2d+2$ dimensional twisted torus with coordinates $\mathbb{X}^I$.

\noindent\textbf{Global Issues}

As in the twisted torus example of section 2, the gauge algebra  only fixes the local structure of the (in this case,
doubled) geometry. This can be seen by the fact that the one forms (\ref{doubled forms}) are invariant under the rigid left action of $\cG$,
which acts on the coordinates infinitesimally as
\begin{eqnarray}
\delta x=\alpha  \qquad \delta\tilde{x}=\tilde{\alpha}-\frac{1}{2}N_{AB}\xi^A\mathbb{X}^B \qquad
\delta\mathbb{X}^A=N^A{}_B\mathbb{X}^B\alpha+\xi^A
\end{eqnarray}
and so the global structure of the doubled group is thus far only determined up to a rigid left action of $\cG$. In general, the doubled space
will be of the form $\cX\simeq\cG/\G$ for some discrete subgroup $\G$. The gauge algebra fixes the local structure of the doubled group, but the
global structure remains undetermined. In particular, the choice of discrete subgroup $\G$ is not determined by the gauge algebra. However,
consistency with the doubled torus picture fixes the identification of most of the coordinates, but not that of $\ti x$. In the case of a
trivial bundle $ \ti x $ is the coordinate for a dual circle with radius inversely related to that of the $x$ circle \cite{Buscher ``A Symmetry
of the String Background Field Equations''}. It seems reasonable to expect that $\G$ should be chosen to be cocompact, so that $\cG/\G$ is
compact. We will return to the discussion of the doubled geometry $\cX\simeq\cG/\G$ and its role in the discussion of T-duality elsewhere, and
show how $\G$  is fixed in particular examples.

\begin{center}
\textbf{Acknowledgment}
\end{center}
RR would like to thank the Institute of Mathematical Sciences at Imperial College London, where this work was initiated, for their hospitality.

\begin{appendix}

\section{$O(d,d)$-Twisted Reduction}

The reduction ansatz is
\begin{eqnarray}
ds^2_{D+1}&=&ds^2_D+\rho\nu^x\otimes\nu^x\nonumber\\
{\cal A}^A(x,y)&=&\left(e^{Nx}\right)^A{}_B\left({\cal A}_{(1)}^B(y)+{\cal A}^B_{(0)}\nu^x\right)\nonumber\\
{\cal B}_{(2)}(x,y)&=&B_{(2)}(y)+B_{(1)}(y)\wedge\nu^x\nonumber\\
{\cal M}^{AB}(x,y)&=&\left(e^{Nx}\right)^A{}_C{\cal M}^{CD}(y)\left(e^{-Nx}\right)_D{}^B\nonumber\\
\phi&=&\varphi+\frac{1}{2}\ln(\rho)
\end{eqnarray}
where the vielbein $\nu^x$ is
\begin{equation}
\nu^x=dx-V^x_{(1)}
\end{equation}
and we have introduced the connection $V^x_{(1)}$ with field strength $F^x_{(2)}=dV^x_{(1)}$.
Using the field redefinitions
\begin{eqnarray}
C_{(2)}&=&B_{(2)}-\frac{1}{2}C_{(1)}\wedge V^x_{(1)}\nonumber\\
C_{(1)}&=&B_{(1)}-\frac{1}{2}L_{AB}{\cal A}^A_{(0)}{\cal A}^B_{(1)}\nonumber\\
C_{(0)}&=&\frac{1}{2}L_{AB}{\cal A}^A_{(0)}{\cal A}^B_{(0)}
\end{eqnarray}
the reduced theory may be written in a manifestly $O(d+1,d+1)$ covariant way
\begin{eqnarray}
{\cal L}_D&=&e^{-\varphi}\left(R*1+*d\varphi\wedge d\varphi+\frac{1}{2}*{\cal H}_{(3)}\wedge {\cal H}_{(3)}+\frac{1}{4}*D{\cal M}_{MN}\wedge
D{\cal M}^{MN}\right. \nonumber\\ &&\left.-\frac{1}{2}{\cal M}_{MN}*{\cal F}^M\wedge{\cal F}^N\right)+V*1
\end{eqnarray}
The scalar potential is now written in the $O(d+1,d+1)$ covariant form
\begin{equation}
V=e^{-\varphi}\left(\frac{1}{4}{\cal M}^{MQ}L^{NT}L^{PS}t_{MNP}t_{QTS}- \frac{1}{12}{\cal M}^{MQ}{\cal M}^{NT}{\cal M}^{PS}t_{MNP}t_{QTS}\right)
\end{equation}

The scalars parameterise the coset $O(d+1,d+1)/O(d+1)\times O(d+1)$
\begin{equation} {\cal M}_{MN}= \left(\begin{array}{ccc}
\rho+{\cal M}_{AB}{\cal A}^A_{(0)}{\cal A}^B_{(0)}+\rho^{-1}C_{(0)}C_{(0)} & \rho^{-1}C_{(0)} & \rho^{-1}C_{(0)}L_{AC}{\cal A}^C_{(0)}+{\cal M}_{AC}{\cal A}^C_{(0)} \\
\rho^{-1}C_{(0)} & \rho^{-1} & \rho^{-1}L_{AC}{\cal A}^C_{(0)}  \\
 \rho^{-1}C_{(0)}L_{BC}{\cal A}^C_{(0)}+{\cal M}_{BC}{\cal A}^C_{(0)}   &  \rho^{-1}L_{AC}{\cal A}^C_{(0)} & {\cal
 M}_{AB}+\rho^{-1}L_{AC}L_{BD}{\cal A}^C_{(0)}{\cal A}^D_{(0)}
\end{array}\right)\nonumber
\end{equation}
The one, two and three-form field strengths are
\begin{eqnarray}\label{Chern-simons field strength}
{\cal H}_{(3)}&=&dC_{(2)}+\frac{1}{2}\left(L_{MN}{\cal A}^M\wedge {\cal F}^N - \frac{2}{3}t_{MNP}{\cal A}^M\wedge {\cal A}^N\wedge {\cal
A}^P\right) \nonumber\\D{\cal M}^{MN}&=&d{\cal M}^{MN}+{\cal M}^{MP}t_{PQ}{}^N{\cal A}^Q+{\cal
M}^{NP}t_{PQ}{}^M{\cal A}^Q\nonumber\\
{\cal F}^M&=&d{\cal A}^M+\frac{1}{2}t_{NP}{}^M{\cal A}^N\wedge{\cal A}^P
\end{eqnarray}
The one-forms form an $O(d+1,d+1)$ vector ${\cal A}^M$ with field strength ${\cal F}^M$
\begin{equation}
{\cal A}^M= \left(\begin{array}{ccc} V^x_{(1)} \\ C_{(1)} \\ {\cal A}^A_{(1)}
\end{array}\right)  \qquad  {\cal F}^M= \left(\begin{array}{cc}
F^x_{(2)} \\ G_{(2)x} \\ {\cal F}^A_{(2)}
\end{array}\right)
\end{equation}
where we have defined
\begin{eqnarray}
G_{(2)x}&=&dC_{(1)}-\frac{1}{2}N_{AB}{\cal A}^A_{(1)}\wedge{\cal A}^B_{(1)}
\end{eqnarray}
Defining $t_{MNP}=L_{MQ}t_{NP}{}^Q$ where $L_{MN}$
  is the  $O(d+1,d+1)$ invariant matrix which takes the block diagonal form
  \begin{equation} L_{MN}=\left(\begin{array}{cc}
L_{xx} & 0 \\
0 & L_{AB} \\
\end{array}\right)  \qquad\qquad    L_{xx}=\left(\begin{array}{cc}
0 & 1 \\
1 & 0 \\
\end{array}\right) \end{equation}
the structure constants are $t _{xB}{}^A=-N^A{}_B$ and $t_{x[AB]}=-N_{AB}$. The presence of $t_{MN}{}^P$ breaks the rigid $O(d+1,d+1)$ symmetry
of the ungauged theory to the subgroup preserving $t_{MN}{}^P$. However, the theory becomes formally invariant under $O(d+1,d+1)$ if the
structure constants are taken to transform covariantly under $O(d+1,d+1)$.

\section{Gauge Symmetry}

In $D+d+1$ dimensions the theory has the antisymmetric tensor transformation symmetry
\begin{equation}
\widehat{B}_{(2)}\rightarrow \widehat{B}_{(2)}+d\widehat{\Lambda}_{(1)}
\end{equation}
The reduction ansatz for the parameter $\widehat{\Lambda}_{(1)}$ on $T^d$ is $\widehat{\Lambda}_{(1)}=\Lambda_{(1)}+\Lambda_{(0)a}\nu^a$. The
remainder of the $U(1)^{2d}$ gauge symmetry comes from the $d$ isometries of the $T^d$, $x^a\rightarrow x^a-\omega^a$, under which $\delta
A^a_{(1)}=-d\omega^a$ and all other fields are invariant. In $D+1$ dimensions this $U(1)^{2d}$ gauge symmetry acts on the fields as
\begin{eqnarray}
\delta_TA^A_{(1)}&=&d\Lambda_{(0)}^A\nonumber\\
\delta_T{\cal B}_{(2)}&=&d\Lambda_{(1)}+\frac{1}{2}L_{AB}\Lambda_{(0)}^A\widehat{{\cal F}}^B_{(2)}
\end{eqnarray}
where we have defined
\begin{eqnarray}
\Lambda^A_{(0)}=\left(%
\begin{array}{c}
  -\omega^a \\
  \lambda_{(0)a} \\
\end{array}%
\right)
\end{eqnarray}

\begin{center}\textbf{Antisymmetric tensor transformations}\end{center}

The duality twist reduction ansatz for the $D+1$ dimensional gauge parameters $\lambda_{(1)}$ and $\lambda^A_{(0)}$ is
\begin{equation}
\Lambda_{(0)}^A=\left(e^{Nx}\right)^A{}_B\lambda^B  \qquad  \Lambda_{(1)}=\lambda_{(1)}+\lambda_x\nu^x
\end{equation}
We denote the infinitesimal variation of the fields under this transformation by $\delta_T$. It is easy to show, by calculating
$d\widehat{\lambda}_{(0)}^A$, that the $D$-dimensional gauge potentials transform as
\begin{eqnarray}
\delta_T{\cal A}^A_{(1)}&=&d\lambda^A+N^A{}_B\lambda^BV^x_{(1)}\nonumber\\
\delta_TC_{(1)}&=&d\lambda_x+N_{AB}\lambda^A{\cal A}^B_{(1)}
\end{eqnarray}

\begin{center}\textbf{$S^1$ Diffeomorphisms}\end{center}

The theory must be invariant under reparameterisations of the circle coordinate
\begin{equation}
x\rightarrow x-\omega
\end{equation}
The matrix $e^{Nx}$ changes as $\left(e^{Nx}\right)^A{}_B\rightarrow
\left(e^{Nx}\right)^A{}_C\left(e^{-N\omega}\right)^C{}_B=\left(e^{Nx}\right)^A{}_C\left(\delta^C{}_B-N^C{}_B\omega+...\right)$. From this is it
easy to see how the $D$-dimensional fields must transform in order for the $D+1$ dimensional ansatz to be invariant. The gauge fields transform
as
\begin{eqnarray}
\delta_Z{\cal A}^A_{(1)}&=&N^A{}_B{\cal A}^B_{(1)}\omega\nonumber\\
\delta_ZV^x_{(1)}&=&-d\omega
\end{eqnarray}

\begin{center}\textbf{Symmetry Algebra}\end{center}

We define
\begin{eqnarray}
\delta_Z=\omega Z_x \qquad  \delta_T=\lambda^AT_A   \qquad  \delta_X=\lambda_xX^x
\end{eqnarray}
where $Z_x$, $X^x$ and $T_A$ are generators of gauge transformations with parameters $\omega$, $\lambda_x$ and $\lambda^A$ respectively. The Lie
algebra of the gauge group is
\begin{eqnarray}
\left[Z_x,T_A\right]&=&-N^B{}_AT_B   \qquad\qquad  \left[T_A,T_B\right]=-N_{AB}X^x
\end{eqnarray}
with all other commutators vanishing.

\end{appendix}

\end{document}